\documentclass[11pt]{article}

\usepackage{algorithm} 
\usepackage{algpseudocode} 
\usepackage{amsmath}
\usepackage{amsfonts}
\usepackage{amssymb}
\usepackage[english]{babel}
\usepackage{booktabs}
\usepackage{caption}
\usepackage{cite}
\usepackage{doi}
\usepackage{enumitem}
\usepackage{float}
\usepackage{graphicx}
\usepackage{hyperref}
\usepackage[capitalise]{cleveref}
\usepackage[utf8]{inputenc}
\usepackage{mathptmx}
\usepackage{microtype}
\usepackage{siunitx}
\usepackage{stfloats}
\usepackage{subfig}
\usepackage{tabularx}
\usepackage{textcomp}
\usepackage{xspace}
\usepackage{color}

\usepackage{./template/arxiv}

\usepackage{./input/ao-math-std}
\usepackage{./input/ao-math-graphs}
\usepackage{./input/ao-math-fields}

\newcommand\perf{\tsb{perf}}
\newcommand\term{\tsb{term}}

\newcommand\etac{\eta\tsb{const}}
\newcommand\orcauth[2]{\href{https://orcid.org/#1}
  {\includegraphics[height=0.7em]{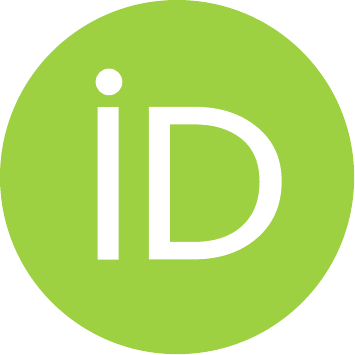}\enspace#2}}

\DeclareSIUnit\mmHg{mmHg}

\title{Branching Exponents of Synthetic Vascular Trees under Different Optimality Principles}
\date{}

\author{
    \orcauth{0000-0003-0276-6029}{Etienne Jessen}\\
	Institute for Mechanics, Computational Mechanics Group\\
	Technical University of Darmstadt\\
	64287 Darmstadt, Germany\\
	\texttt{etienne.jessen@tu-darmstadt.de} \\
\And
    \orcauth{0000-0002-6343-9809}{Marc C. Steinbach} \\
	Institute of Applied Mathematics\\
	Leibniz Universität Hannover\\
	30167 Hannover, Germany\\
	\texttt{mcs@ifam.uni-hannover.de} \\
\And
    \orcauth{0000-0003-1962-238X}{Charlotte Debbaut} \\
	IBiTech – Biommeda \\
	Ghent University\\
	Ghent, Belgium \\
	\texttt{Charlotte.Debbaut@UGent.be} \\
\And
    \orcauth{0000-0002-9068-6311}{Dominik Schillinger} \\
	Institute for Mechanics, Computational Mechanics Group\\
	Technical University of Darmstadt\\
	64287 Darmstadt, Germany\\
	\texttt{dominik.schillinger@tu-darmstadt.de} \\
}

\begin{document}
\maketitle

\begin{abstract}
\textit{Objective:} The branching behavior of vascular trees is often characterized using Murray's law. We investigate its validity using synthetic vascular trees generated under global optimization criteria.
\textit{Methods:} Our synthetic tree model does not incorporate Murray's law explicitly.
  Instead, we assume it holds implicitly and investigate the effects of different physical constraints and optimization goals on the branching exponent that is now allowed to vary locally.
  In particular, we include variable blood viscosity due to the Fåhræus--Lindqvist effect and enforce an equal pressure drop between inflow and the micro-circulation.
  Using our global optimization framework, we generate vascular trees with over one million terminal vessels and compare them against a detailed corrosion cast of the portal venous tree of a human liver.
\textit{Results:} Murray's law is implicitly fulfilled when no additional constraints are enforced, indicating its validity in this setting.
  Variable blood viscosity or equal pressure drop leads to deviations from this optimum, but with the branching exponent inside the experimentally predicted range between 2.0 and 3.0.
  The validation against the corrosion cast shows good agreement from the portal vein down to the venules.
\textit{Conclusion:} Not enforcing Murray's law explicitly reduces the computational cost and increases the predictive capabilities of synthetic vascular trees.
\textit{Significance:} The ability to study optimal branching exponents across different scales can improve the functional assessment of organs
\end{abstract}


\newpage
\section{Introduction}
\label{sec:introduction}
The cardiovascular system is responsible for transporting blood to and from all cells in the human body, leading to hierarchical networks of vessels, called vascular trees, inside each organ.
According to Murray \cite{murray1926physiological},
this hierarchy obeys scaling relations based on the minimization of the total energy expenditure of the system.
Many factors influence and constrain this minimization process, such as the type and shape of the organ supplied, the demand for the organ's cells, and the existence of vascular diseases.
The goals and constraints guiding their structural development and influence on the vascular system have yet to be entirely understood, even though extensive work has been carried out for over a century.
Thus the analysis of vascular diseases based on the anatomy and physiology of the vascular structure remains a challenge.

Murray first described a minimization problem for vascular segments in 1926 \cite{murray1926physiological, murray1926physiological2}.
Here, a tree is approximated as a bifurcating network consisting of rigid tubes, and the physical principles for fluid flow follow Poiseuille's law.
The goal is to minimize the total power of the tree network with its minimum characterized by \emph{Murray's law}.
It describes the relationship of the radius of a parent vessel $r_0$ against the radii of its children's vessels $\left( r_1, r_2\right)$ as a power law with
\begin{align}\label{eq:murrays-law}
    r_0^\gamma = r_1^\gamma + r_2^\gamma.
\end{align}

The branching exponent $\gamma$ became an essential parameter for characterizing the branching behavior of vascular trees.
In Murray's original formulation, $\gamma = 3.0$ is constant across the entire network.

An extensive number of studies have been conducted to investigate Murray's law experimentally \cite{horsfield1989diameters, jiang1994diameter, vanbavel1992branching}.
In general, exponents between $2.0$ and $3.0$ were measured.
In \cite{nakamura2014radius}, exponents were observed even going over the theoretical limit of 3.0 with $\gamma = 3.2$.
Multiple theoretical studies have analyzed the possible factors contributing to these branching behaviors.
An extension to Murray's law was proposed by Uylings \cite{uylings1977optimization}, which incorporated the effects of turbulent flow into the minimization problem.
Results show branching exponents as low as $2.33$ for turbulent flow.
In \cite{painter2006pulsatile}, a vascular model was investigated, which considered the role of elastic tubes.
Compared to rigid tubes, the effect of pulsatile flow lowered the optimal value to $2.3$.
Zhou, Kassab, and Molloi \cite{zhou1999design, kassab2006scaling} generalized Murray's law hypothesis to an entire coronary arterial tree by defining a vessel segment as a stem and the tree distal to the stem as a crown.
They showed that $\gamma$ deviates from $3.0$ even for steady-state flow and depends on the ratio between metabolic demand and viscous power dissipation.

An alternative approach to investigate branching exponents is to construct vascular trees synthetically.
The most well-known generation method here is constrained constructive optimization (CCO) \cite{schreiner1993computer}.
The local optimization approach is directly based on Murray's minimization principles and allows to investigate different, albeit constant, values for $\gamma$, like $2.55$ \cite{karch1999three} or $3.0$ \cite{schreiner2003heterogeneous}.
Another approach, based on Simulated Annealing (SA), included the branching exponent as an optimization parameter \cite{keelan2021role}.
Results show that the vascular topology and the metabolic demand significantly influence the value of the branching exponent.
Recently, the authors extended the CCO approach to finding a synthetic tree optimal both in (global) geometry and topology \cite{jessen2022rigorous}.
Finding the optimal geometry is cast into a nonlinear optimization problem (NLP), which allows the investigation of various possible goal functions and constraints.

In this paper, we utilize this flexibility and go beyond previous studies by allowing the branching exponent $\gamma$ to vary locally.
Furthermore, we include a blood viscosity law based on the Fåhræus--Lindqvist effect \cite{pries1994resistance} and enforce equal pressure drop to terminal vessels.
The goal is to investigate the change in branching exponents under these influences.
We start by introducing the relevant definitions and assumptions to generate synthetic trees.
We then cast our goals and constraints into NLPs and introduce our optimization framework in more detail.
Finally, we generate full portal venous trees of the human liver with up to one million terminal vessels and compare them against a vascular corrosion cast of a human liver \cite{debbaut2010vascular, debbaut2014analyzing}.

\section{Methods}
\subsection{Definitions and assumptions}
\begin{figure}[ht]
  \centering
  \includegraphics[trim=20 0 0 0, clip=true,width=0.36\textwidth]{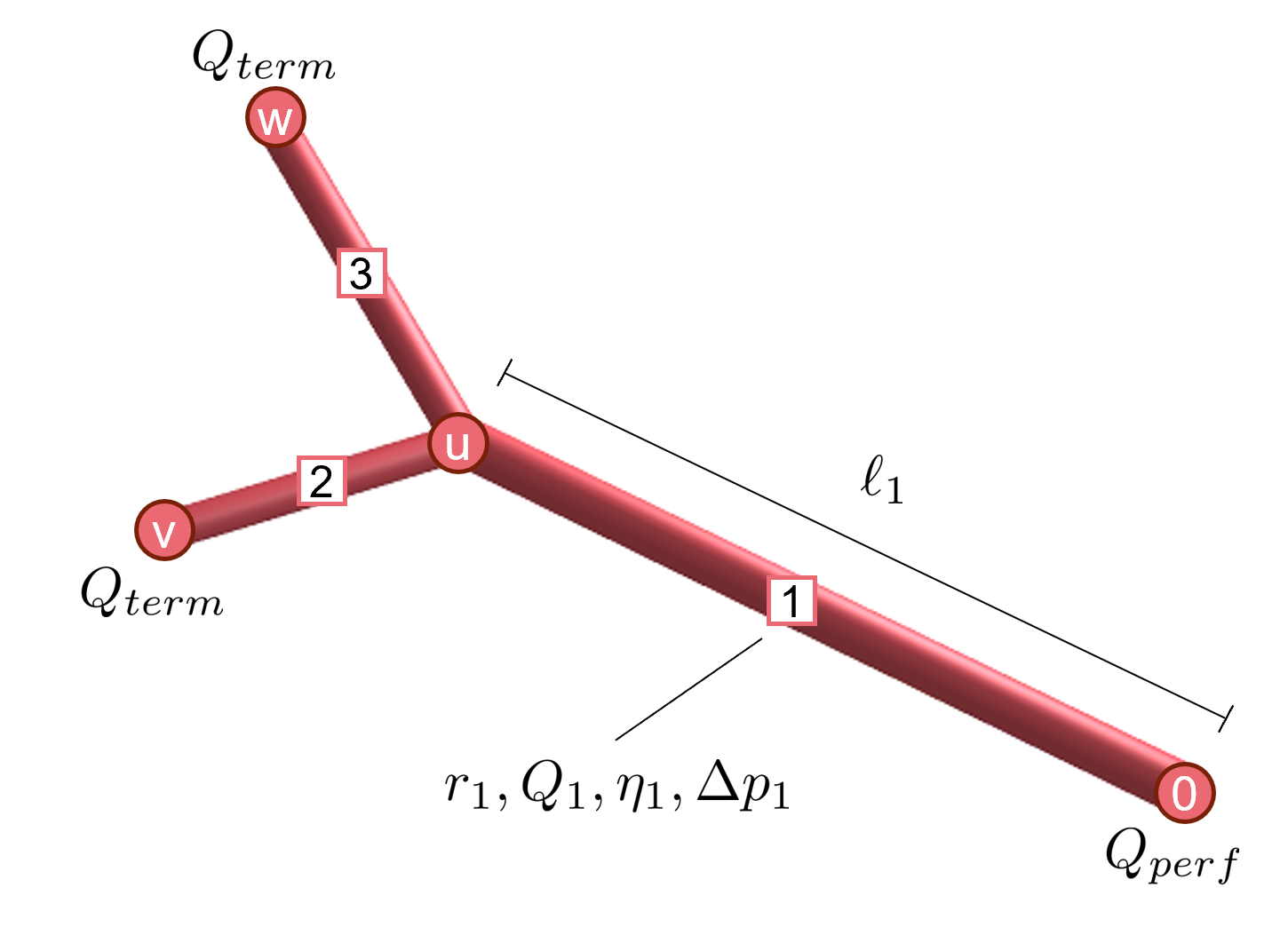}
  \caption{Schematic of a vascular tree and its relation to nodes and segments. Red circles denote a node, and white rectangles a segment. This tree has a given inflow $Q_1 = Q\perf$ and equal terminal outflow $Q_2 = Q_3 = Q\term$ through each of the outlets (leaves)}
  \label{fig:vascular_graph_structure}
\end{figure}

We represent a vascular tree as a directed branching network $\Tree = (\Nodes, \Arcs)$ with nodes $u \in \Nodes$ and segments $a \in \Arcs$.
Each segment $a = uv$ connects a proximal node $x_u$ with a distal node $x_v$.
It approximates a vessel as a rigid and straight cylindrical tube and is defined by its radius $r_a$, length $\ell_a = \norm{x_u - x_v}$, volumetric flow $Q_a$ and apparent blood viscosity $\eta_a$.
The distal nodes of \emph{terminal segments} are terminal nodes (\emph{leaves}) $v \in \Leaves$, and the proximal node of the (single) \emph{root segment} is the root node $x_0$.
A synthetic vascular tree perfuses blood at a steady state from the root segment down to the terminal segments inside a given (non-convex) perfusion volume $\Omega \subset \R^3$, schematically shown in \cref{fig:vascular_graph_structure}

As in Murray's original paper \cite{murray1926physiological}, we assume laminar flow and approximate blood as an incompressible homogeneous Newtonian fluid.
We express the hydrodynamic resistance $R_a$ of segment $a$ by Poiseuille's law with
\begin{equation}
  R_a = \frac{8 \eta_a}{\pi} \frac{\ell_a}{r_a^4} \quad\forall a \in \Arcs.
\end{equation}
The pressure drop over a segment can now be computed as
\begin{equation}
  \Delta p_a = R_a Q_a \quad\forall a \in \Arcs,
\end{equation}
and the pressure at a node $v$ follows with
\begin{equation}
    p_v = p_u - \Delta p_a \quad\forall uv \in \Arcs.
\end{equation}

We further assume that the (known) perfusion flow $Q\perf$ is homogeneously distributed among all $N$ terminal segments, leading to a terminal flow value $Q\term = Q\perf/N$.
All remaining flow values can then be computed using Kirchhoff's law with $Q_{uv} = \sum_{vw \in \Arcs}  Q_{vw} \forall v \in \Nodes \setminus \left({0} \cup \Leaves\right)$.

\begin{figure}[ht]
  \centering
  \includegraphics[trim=15 0 0 0, clip=true,width=0.48\textwidth]{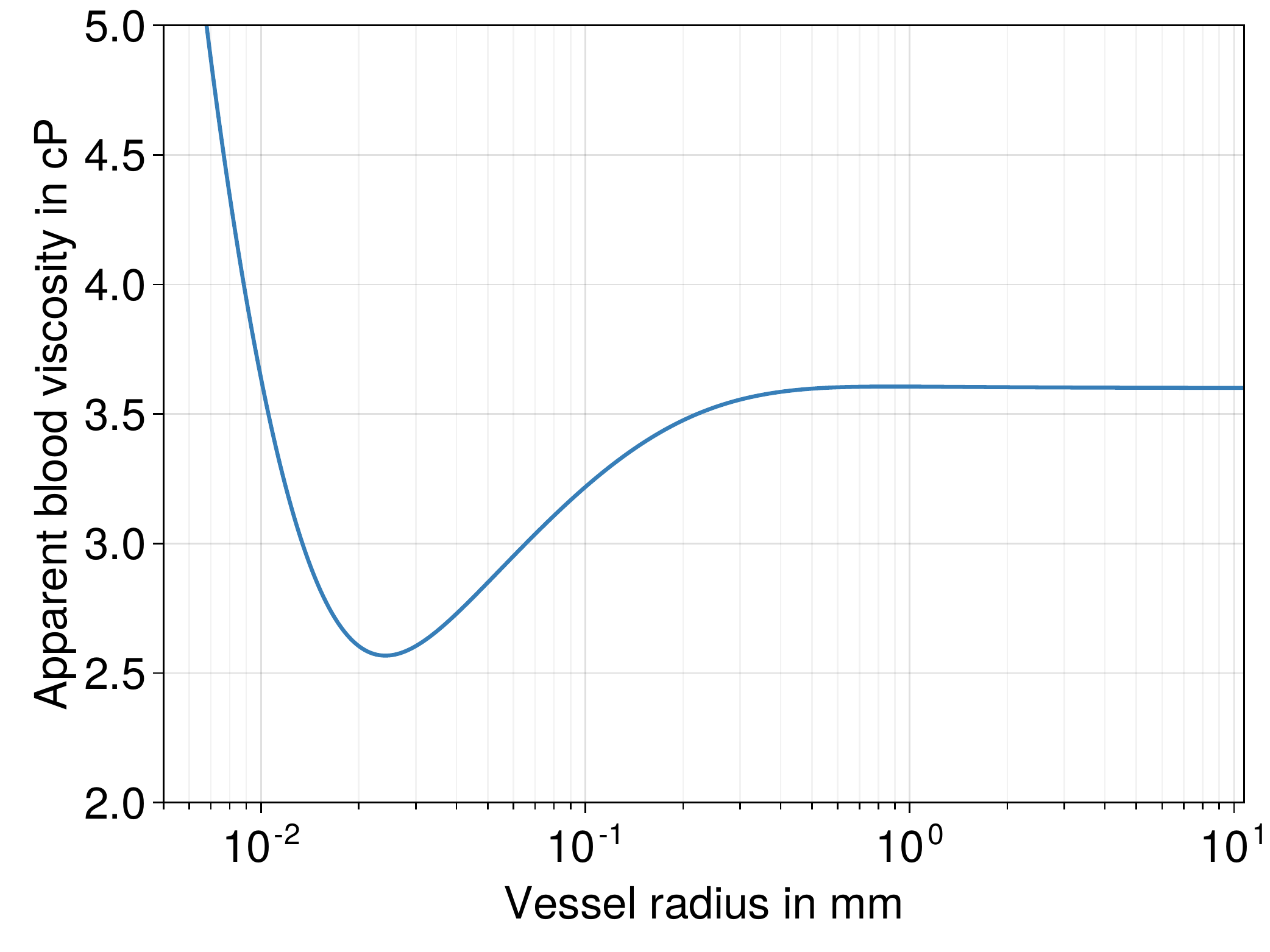}
  \caption{Change in apparent blood viscosity due to the Fåhræus--Lindqvist effect as approximated by Pries et al. \cite{pries1994resistance}}
  \label{fig:var_eta_comparison}
\end{figure}

We aim at generating vessels down to the smallest arterioles/venules with typical radii in the range of \SI{0.015}{mm} to \SI{0.1}{mm}.
The Fåhræus--Lindqvist effect \cite{fahraeus1931viscosity} should be accounted for at this scale.
It describes how the blood viscosity decreases as the vessel diameter decreases.
The tendency of red blood cells to migrate toward the vessel center is largely responsible for this effect.
In turn, this forces plasma toward the walls and decreases peripheral friction.
At the smallest vessels with radii approaching the radii of red blood cells, the viscosity sharply rises again.
Pries et al. \cite{pries1994resistance} derived an empirical relationship for this behavior with
\begin{align}\label{eq:var-eta-pries}
    \eta(r_a) &= \eta_p \bigl( \kappa + \kappa^2 \bigl(\eta_{45} - 1\bigr) \bigr),\\
    \eta_{45} &= 6 \exp(-170r_a / \text{mm}) - \\
              &2.44 \exp(-8.08 (r_a / \text{mm})^{0.645}) + 3.2,\\
  \kappa &= \frac{r_a^2}{(r_a - 0.00055\text{mm})^2},
\end{align}
where $\eta_p$ is the viscosity of the plasma, which we set to $\eta_p = \SI{1.125}{cP}$.
$\eta_{45}$ is the relative apparent blood viscosity for a discharge hematocrit of \num{0.45}.
This relationship is depicted in \cref{fig:var_eta_comparison} for the relevant radii between \SI{0.015}{mm} and \SI{10}{mm}.

\subsection{Design goals and constraints}
\subsubsection{Murray's minimization problem}
The original minimization formulation by Murray states that the total power of a vascular tree consists of the metabolic power required to sustain blood $P\tsb{vol}$ and the viscous power $P\tsb{vis}$ required to pump blood from the root down to the micro-circulation.
The cost function of a tree is then defined as
\begin{align}
    f_\mathbb{T} = P\tsb{vol} + P\tsb{vis} = \sum_{a \in \mathbb{A}} m_b\pi\ell_a r_a^2 +\frac{8 \eta_a}{\pi} \frac{\ell_a}{r_a^4}Q_a^2,
\end{align}
where $m_b$ is the metabolic demand of blood in \si{\uW\per\cubic\mm}.
As described in \cite{jessen2022rigorous}, we can now include the nodal positions $x$, the length $\ell$ and radii $r$ as well as the blood viscosity $\eta$ in the vector of optimization variables $y$, leading to $y_1 = (x, \ell, r, \eta)$.
We add physical lower bounds $\ell^-,\ r^-$ and $\eta^-$ and, for numerical efficiency, upper bounds $\ell^+,\ r^+$ and $\eta^+$.
The best geometry is then found in the rectangle defined as
\begin{align}
    Y_1 = \mathbb{R}^{3|\mathbb{V}|} \times [\ell^-, \ell^+]^\mathbb{A} \times [r^-, r^+]^\mathbb{A} \times [\eta^-, \eta^+]^\mathbb{A}.
\end{align}

Our NLP "\emph{Power minimization}" finally reads:
\begin{align}\label{nlp-murray}
  \min_{y_1 \in Y_1} \quad
  & \sum_{a \in \Arcs} m_b\pi\ell_a r_a^2 + 8\eta/\pi Q_a^2\ell_a/r_a^4\\
  \stq
  \label{eq:nlp1-fix-x}
  &0 = x_u - \_x_u, & u &\in \Nodes_0 \cup \Leaves\\
  \label{eq:nlp1-length}
  &0 = \ell_{uv}^2 - \norm{x_u - x_v}^2, & uv &\in \Arcs\\
  \label{eq:nlp1-var-eta}
  &0 = \eta_a - \eta(r_a) & a &\in \Arcs
\end{align}

\cref{eq:nlp1-fix-x} fixes the position of terminal nodes and \cref{eq:nlp1-length} ensures consistency between nodal positions and segment length.
The third constraint in \cref{eq:nlp1-var-eta} enforces the Fåhræus--Lindqvist effect as defined in \cref{eq:var-eta-pries}.

\subsubsection{Enforcing equal pressure drop}
In Murray's original formulation, no consideration of the resulting pressure values at terminal segments was given. This terminal pressure is a crucial parameter for the regulation of blood flow and blood velocity at the microcirculatory domains. Since we assume these domains are roughly homogeneous across the organ and have equal demand, the pressure should not differ significantly.
Therefore we enforce equal pressure at each terminal segment by adding the pressure $p_v$ at each node as a new unknown in our (second) NLP with variables $y_2 = (x, \ell, r, \eta, p)$, leading to \begin{align}
    Y_2 = \mathbb{R}^{4|\mathbb{V}|} \times [\ell^-, \ell^+]^\mathbb{A} \times [r^-, r^+]^\mathbb{A} \times [\eta^-, \eta^+]^\mathbb{A}.
\end{align}

Secondly, we constrain the pressure drop between the root and the terminal nodes as a prescribed constant value $\Delta_p$.
Since the viscous power at each segment $a$ is directly proportional to the pressure drop by a factor of $Q_a$, the total viscous power $P\tsb{vis}$ becomes a constant.
Thus, we can remove it from the cost function.
Finally, we can drop the constant factor $m_b$, leading to a minimization goal proportional to the tree volume $V_\Tree$.
This formulation is used in most synthetic tree studies, e.g., \cite{schreiner1993computer, karch1999three, kretowski2003physiologically, keelan2021role, jessen2022rigorous}.
Our NLP "\emph{Volume minimization}" then reads:
\begin{align}\label{nlp-eq-pressure}
  \min_{y_2 \in Y_2} \quad
  & \sum_{a \in \Arcs} \pi\ell_a r_a^2\\
  \stq
  \label{eq:nlp2-fix-x}
  &0 = x_u - \_x_u, & u &\in \Nodes_0 \cup \Leaves\\
  \label{eq:nlp2-length}
  &0 = \ell_{uv}^2 - \norm{x_u - x_v}^2, & uv &\in \Arcs \\
  \label{eq:nlp2-deltap1}
  &0 =  p_u - p_v - (8 \eta / \pi) Q_{uv} \ell_{uv} / r_{uv}^4,  & uv &\in \Arcs\\
  \label{eq:nlp2-deltap2}
  &0 = p_u, & u &\in \Leaves\\
  \label{eq:nlp2-deltap3}
  &0 = p_0 - \Delta_p,\\
  \label{eq:nlp2-var-eta}
  & 0 = \eta_a - \eta\left(r_a\right) & a &\in \Arcs
\end{align}

\textbf{Remark 1:}
Murray's law, as stated in \cref{eq:murrays-law}, is not incorporated explicitly.
Instead, we assume it holds implicitly and compute the associated branching exponent $\gamma$ using the Newton-Raphson method after the optimization is finished.

\subsubsection{Additional optimization variants}\label{sec:NLP-variants}
To better isolate the individual influence of different factors, we define additional variants of our two minimization problems.
Firstly, we simplify both problems to a constant apparent viscosity $\etac = \SI{3.6}{cP}$, removing $\eta$ from the vector of optimization variables and dropping the corresponding constraints \cref{eq:nlp1-var-eta} and \cref{eq:nlp2-var-eta}.
Secondly, we investigate the influence of the metabolic demand $m_b$ and the total pressure drop $\Delta_p$.
We consider values between \SI{0.1}{\uW\per\cubic\mm} and \SI{1.0}{\uW\per\cubic\mm} for the metabolic demand, but note that estimates of this parameter vary significantly \cite{liu2007vascular}.
For the total pressure drop, we set the terminal pressure to $p\term = \SI{6}{\mmHg}$ and vary the root pressure between \SI{10}{\mmHg} and \SI{14}{\mmHg}.
Finally, for each variant, a separate tree is generated, where Murray's law (see \cref{eq:murrays-law}) is enforced directly with a single exponent $\gamma\tsb{opt}$, included in the optimization variables.
All variants include the same root flow $Q\perf = \SI{1.1}{l/min}$.

\subsection{Generation framework}
We generate our synthetic trees using the framework introduced in \cite{jessen2022rigorous}, which we summarize in the following.
First, we generate $N\tsb{topo}$ terminal nodes on a regular cubic grid inside our organ's volume.
The root position is manually set and connected to the geometric center of the volume, which in turn is connected to all terminal nodes.
We swap segments to explore new topologies from this initial (fan-shaped) tree.
A \emph{swap} detaches a segment from its parent and connects it with another segment.
After each swap, the geometry is optimized by solving the corresponding NLP.
The newly created topology is accepted on the basis of an SA approach.
After topology optimization, we grow the tree using a modified CCO approach.
Here, we optimize the global geometry each time after adding $N\tsb{geo}$ new terminal nodes and then increase $N\tsb{geo}$ heuristically based on the current density of the tree.
Notably, we drop the local optimization of branching positions and set them to their flow-weighted mean, similar to \cite{guy20193d}.
Due to our repeated global geometry optimization, this simplification had no significant impact on the final tree structure.
In the last step of the optimization, we delete all segments that reached the lower bound $\ell^-$ (\emph{degenerate segments}),  possibly creating $n$-furcations $\left(n\geq3\right)$.

We then classify the hierarchy throughout the finished tree by assigning each segment an order number corresponding to the Strahler ordering method \cite{strahler1957quantitative}.
Continuous segments of the same Strahler order correspond to one \emph{vessel}.
Additionally, we employ an ordering scheme based on \cite{jiang1994diameter} to allow direct comparison (in reverse order) to the \emph{generation} notation used for the vascular corrosion cast in \cite{debbaut2014analyzing}.

\textbf{Remark 2:}
The complete optimization framework is only applied once to obtain a common topology.
We solve each NLP variant with this topology to get the corresponding global geometry.
We choose this method to focus on the geometry changes and to allow a direct comparison of branching exponents and radii at the same branch types.

\section{Results}
\subsection{Overall structure of the synthetic portal vein}
\begin{figure*}[ht]
  \centering
  \includegraphics[trim=11 0 0 0, clip=true,width=1.0\textwidth]{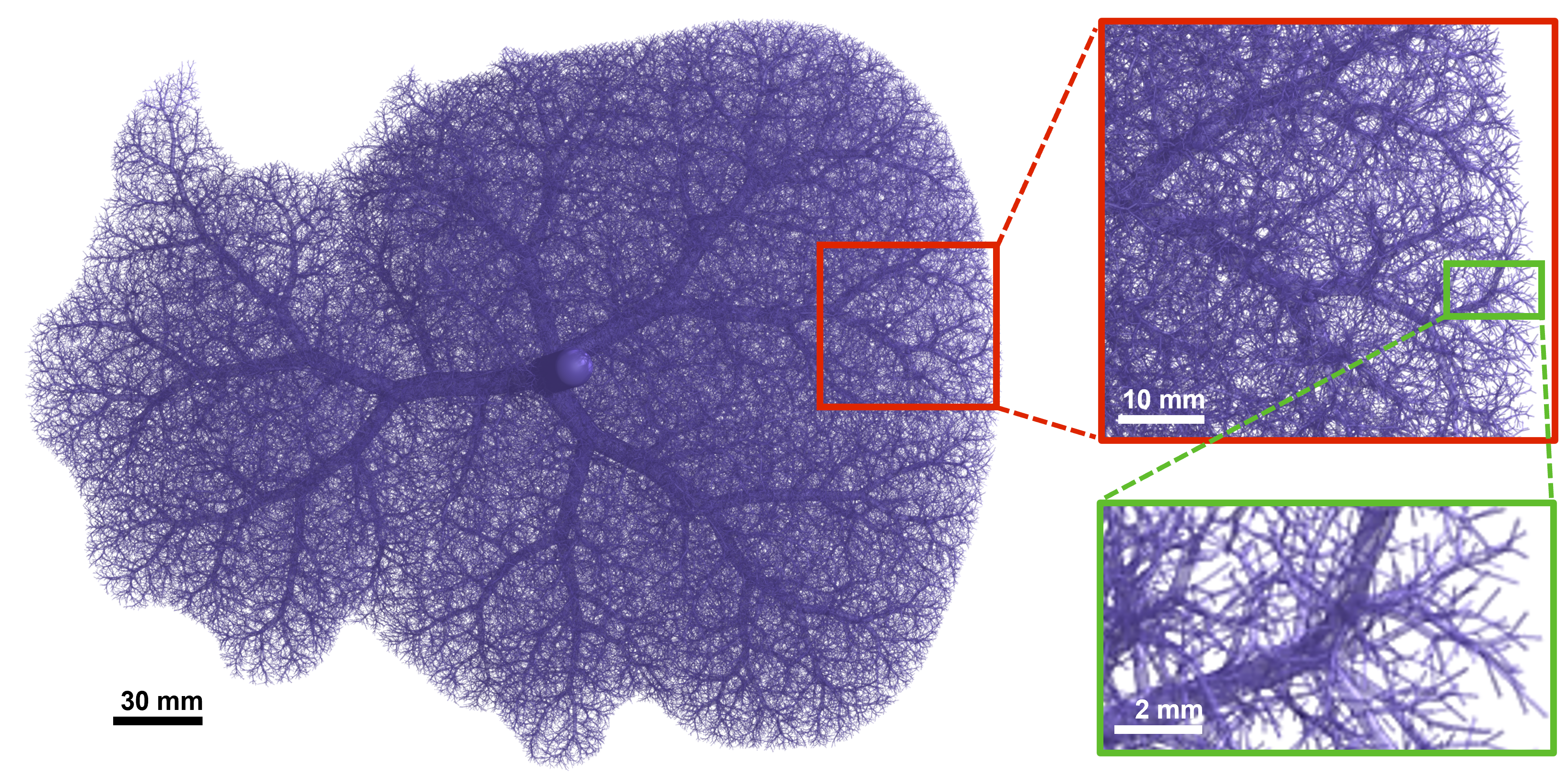}
  \caption{Complete synthetic vascular tree of the portal vein of a human liver with $1,000,000$ terminal vessels (volume minimization with variable viscosity and $p\tsb{root} = \SI{10}{\mmHg}$).
            Two zoom levels highlight the hierarchical structure at different scales.
            The radii are between $\SI{5.1}{mm}$ (root vessel) and $\SI{0.017}{mm}$ (smallest terminal vessel)}
  \label{fig:complete_tree}
\end{figure*}

The full synthetic portal vein tree for the case of volume minimization with variable viscosity is depicted in \cref{fig:complete_tree}.
The left side shows the complete tree inside the non-convex liver domain with two zoom levels.
The tree splits into four major branches, which further split into $8$ main branches.
These results align with previous results of a sparser tree in \cite{jessen2022rigorous}.

For a detailed comparison between the different optimization variants discussed in \cref{sec:NLP-variants}, we summarized the results in \cref{tab:results_overview}.
Here, a column represents the results of a single variant, with the first four columns corresponding to the NLP "Power minimization" and the last four columns corresponding to the NLP "Volume minimization".
For "Power minimization", we included the results for metabolic demands $m_b$ of \SI{0.1}{\uW\per\cubic\mm} and \SI{1.0}{\uW\per\cubic\mm}.
Similarly, for "Volume minimization", we included the results for root pressures $p\tsb{root}$ of \SI{10}{\mmHg} and \SI{14}{\mmHg}.
The last row indicates the results of each variant after enforcing a single (optimal) branching exponent $\gamma\tsb{opt}$.

For "Power minimization", an increase in metabolic demand $m_b$ from \SI{0.1}{\uW\per\cubic\mm} to \SI{1.0}{\uW\per\cubic\mm} leads to an increase in viscous power $P\tsb{vis}$, shown in the first row of \cref{tab:results_overview}, by around \SI{364}{\percent} and a reduction in volume $V_\mathbb{T}$, shown in the second row \cref{tab:results_overview}, by around \SI{53}{\percent}.
Similarly, for the "Volume minimization", an increase in the root pressure from \SI{10}{\mmHg} to \SI{14}{\mmHg} leads to a \SI{320}{\percent} increase in viscous power and a \SI{51}{\percent} reduction in volume.
The Fåhræus--Lindqvist effect had a minor influence. It decreased the viscous power by around \SI{1}{\percent} and the total volume by around \SI{4}{\percent} in all cases.

\subsection{Vessel radii}
The root radius $r\tsb{root}$ is shown in the third row of \cref{tab:results_overview}. It decreased by \SI{31}{\percent} after the metabolic demand $m_b$ was increased for "Power minimization".
Similarly, in the case of "Volume minimization", it decreased by \SI{30}{\percent} after the root pressure $p\tsb{root}$ was increased to \SI{14}{\mmHg}.
In all cases, the Fåhræus--Lindqvist effect had a limited influence on the root radius with changes less than \SI{0.1}{\percent}.
However, a significant decrease in radius can be observed for vessels between Strahler order $1$ and $6$, shown in \cref{fig:radius_comparison}.
In both NLP cases, this decrease was highest at the terminal vessels with \SI{2}{\percent} for "Power minimization" (\cref{fig:radius_min_power}) and \SI{1.5}{\percent} for "Volume minimization" (\cref{fig:radius_min_volume}).
A notable difference between both cases is the variance of radii at each Strahler order.
In the case of "Power minimization", the highest variance is observed at Strahler order $6$, whereas the terminal radii are constant.
In the case of "Volume minimization", the highest variance is at the terminal segments and decreases as the Strahler order increases.

\bgroup
\newcommand\?{\rlap*}
\newcommand\cl{\cline}
\newcommand\mc[1]{\multicolumn{#1}{c}}
\newcommand\I{$\vphantom{f_f^{f^f}}$}
\renewcommand\arraystretch{1.15}
\begin{table*}[ht]
  \centering
  \caption{Comparison of the results between the different variants of
    our two minimization problems as introduced in \cref{sec:NLP-variants}.
    The branching exponents $\gamma\tsb{opt}$ in the last row
    (marked with *) are the results of separate runs for each variant,
    where a single constant branching exponent was enforced.}
  \begin{tabular*}{\textwidth}{l*8{@{\extracolsep{\fill}}r}}
    \toprule
    & \mc4{Power minimization} & \mc4{Volume minimization} \\
    \cl{2-5} \cl{6-9}
    & \mc2{$m_b = \SI{0.1}{\uW\per\cubic\mm}$}
    & \mc2{$m_b = \SI{1.0}{\uW\per\cubic\mm}$}
    & \mc2{$p\tsb{root} = \SI{10.0}{\mmHg}$}
    & \mc2{$p\tsb{root} = \SI{14.0}{\mmHg}$}\I \\
    \cl{2-3} \cl{4-5} \cl{6-7} \cl{8-9}
    Parameter
    & \mc1{$\etac$} & \mc1{$\eta(r)$} & \mc1{$\etac$} & \mc1{$\eta(r)$}
    & \mc1{$\etac$} & \mc1{$\eta(r)$} & \mc1{$\etac$} & \mc1{$\eta(r)$}\I \\
    \cl{1-1}
    \cl{2-2} \cl{3-3} \cl{4-4} \cl{5-5} \cl{6-6} \cl{7-7} \cl{8-8} \cl{9-9}
    $P\tsb{vis}$ in \si{mW}
    & 2.67 & 2.65 & 12.34 & 12.23 & 3.33 & 3.33 & 13.99 & 13.99\I \\
    $V_\Tree$ in \si{mm^3}
    & 53,400.60 & 52,283.94 & 24,786.36 & 24,042.08
    & 50,413.84 & 49,158.58 & 24,587.02 & 23,154.13 \\
    $r\tsb{root}$ in \si{mm}
    & 5.35 & 5.35 & 3.64 & 3.63 & 5.10 & 5.09 & 3.56 & 3.54 \\
    $p\term$ in \si{mm Hg}
    & $[10.34, 11.68]$ & $[10.41, 11.73]$ & $[1.15, 10.67]$ & $[1.74, 10.98]$
    & 6.00 & 6.00 & 6.00 & 6.00 \\
    $\gamma$
    & 3.00 & $[2.90, 3.00]$ & 3.00 & $[2.90, 3.00]$
    & $[1.75, 3.01]$ & $[1.76, 3.17]$ & $[1.43, 3.00]$ & $[1.46, 3.08]$ \\
    $\gamma\tsb{opt}$ (constant)
    & 3.00\? & 2.91\? & 3.00\? & 2.92\? & 2.84\? & 2.76\? & 2.82\? & 2.74\? \\
    \bottomrule
  \end{tabular*}
  \label{tab:results_overview}
\end{table*}

\begin{figure*}[ht]
  \centering
  \subfloat[Power minimization ($m_b = \SI{0.1}{\uW\per\cubic\mm}$)]
  {\includegraphics[trim=0 0 0 0,clip=true,width=0.49\textwidth]
    {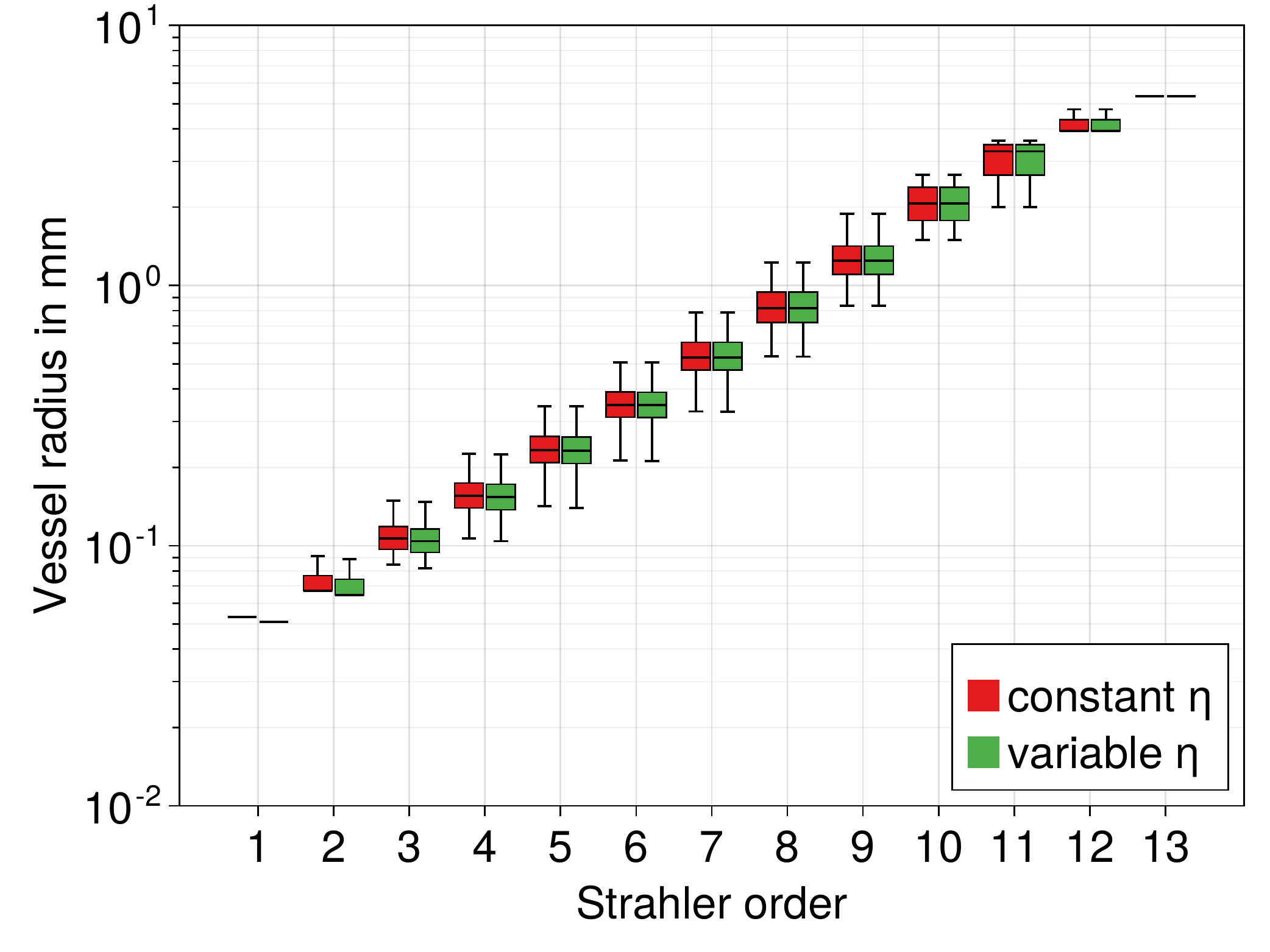}%
    \label{fig:radius_min_power}}\hfill
  \subfloat[Volume minimization ($p\tsb{root} = \SI{10}{\mmHg}$)]
  {\includegraphics[trim=0 0 0 0,clip=true,width=0.49\textwidth]
    {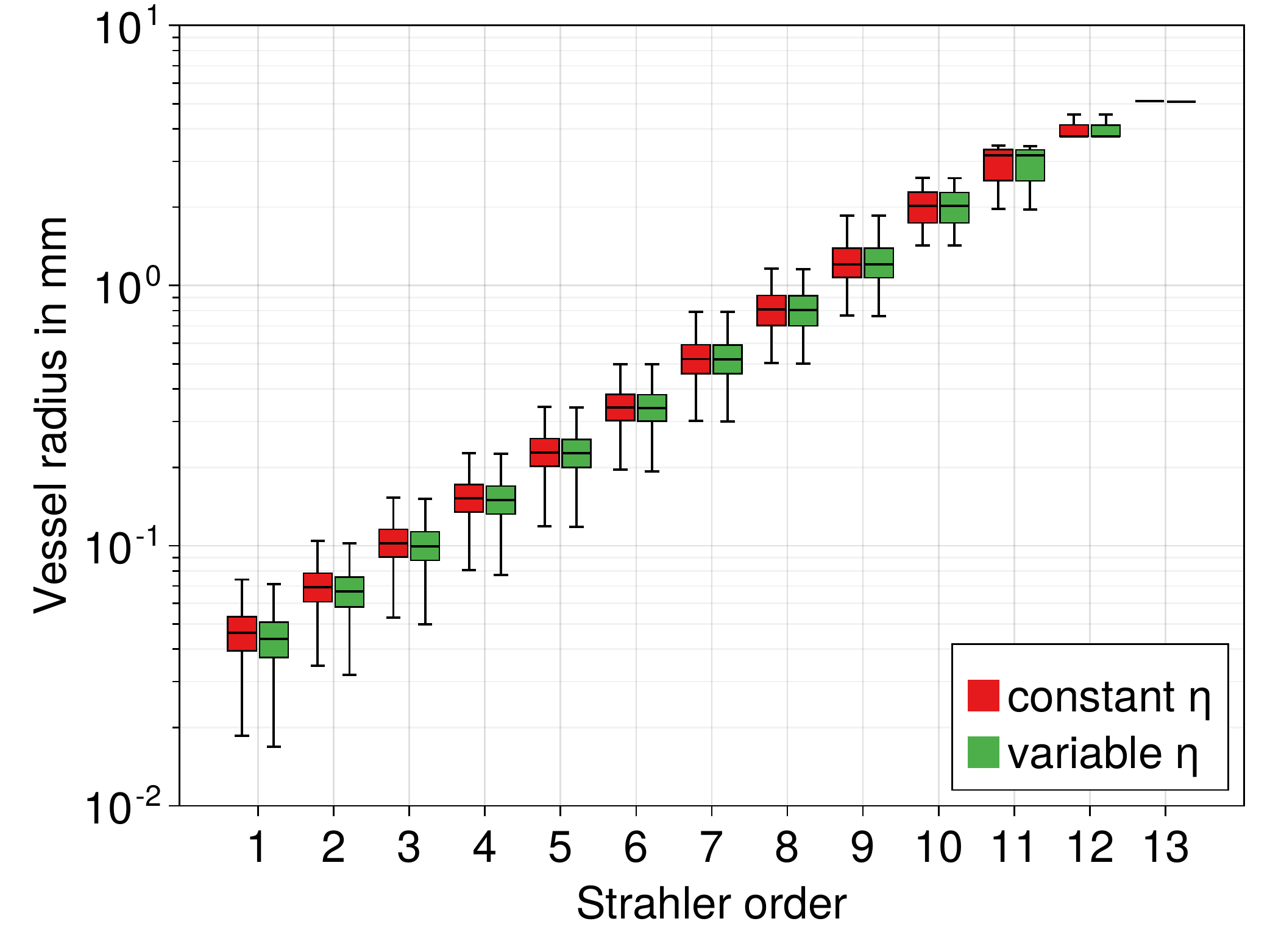}%
    \label{fig:radius_min_volume}}
  \caption{Influence of Fåhræus--Lindqvist effect
    on vessel radii for different Strahler orders}
  \label{fig:radius_comparison}
\end{figure*}

\subsection{Pressure drop}
After "Power minimization", the terminal pressures are not constant across the tree but exhibit a wide range of values, see \cref{tab:results_overview} row 4 column 1--4.
This range widens further for higher metabolic demands and also increases the mean total pressure drop from the root to the terminal segments, shown in \cref{fig:pressure-drop-density}.
In \cref{fig:pressure_min_power}, the pressure values at different Strahler orders are shown for $m_b = \SI{0.1}{\uW\per\cubic\mm}$.
Pressure values drop and variances increase with decreasing Strahler order.
Including the Fåhræus--Lindqvist effect leads to slightly higher pressure values for the Strahler orders $1$ to $6$.
The terminal pressures after "Volume minimization" are fixed to $p\term = \SI{6}{\mmHg}$ as enforced by \cref{eq:nlp2-deltap1} -- \cref{eq:nlp2-deltap3}.
The effect of these constraints is highlighted in \cref{fig:pressure_min_volume} for root pressure $p\tsb{root} = \SI{10}{\mmHg}$.
In contrast to "Power minimization", variances are significantly higher at the intermediate Strahler orders $3$ to $11$.
Furthermore, the influence of the Fåhræus--Lindqvist effect is more pronounced, decreasing the pressure values between Strahler order $2$ to $10$.
\begin{figure}[ht]
  \centering
  \includegraphics[trim=0 0 0 0, clip=true,width=0.48\textwidth]{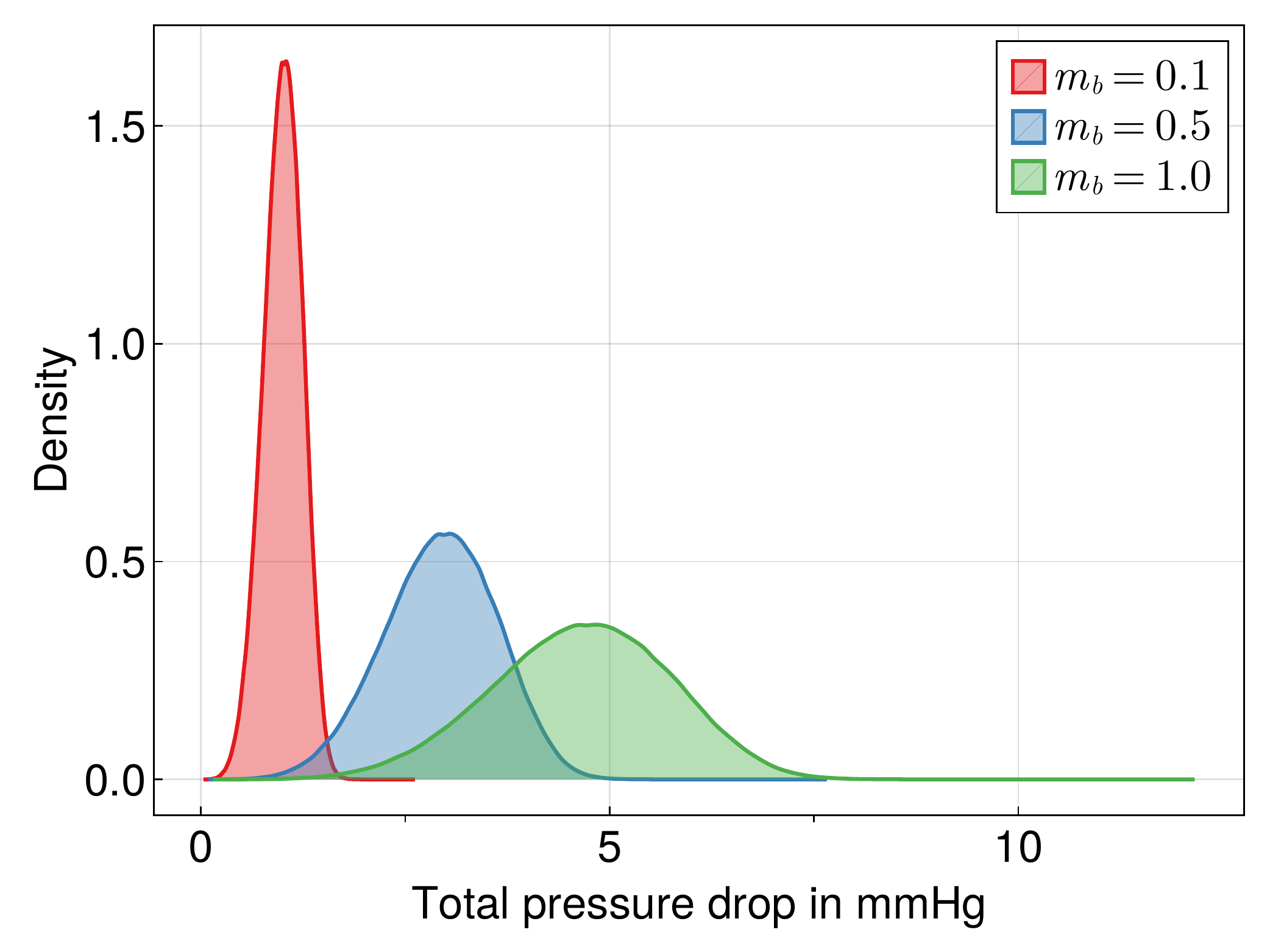}
  \caption{Density plot of the total pressure drop from root vessel to terminal vessels for the power minimization under different metabolic demand}
  \label{fig:pressure-drop-density}
\end{figure}

\begin{figure*}[ht]
  \centering
  \subfloat[Power minimization ($m_b = \SI{0.1}{\uW\per\cubic\mm}$)]
  {\includegraphics[trim=0 0 0 0,clip=true,width=0.5\textwidth]
  {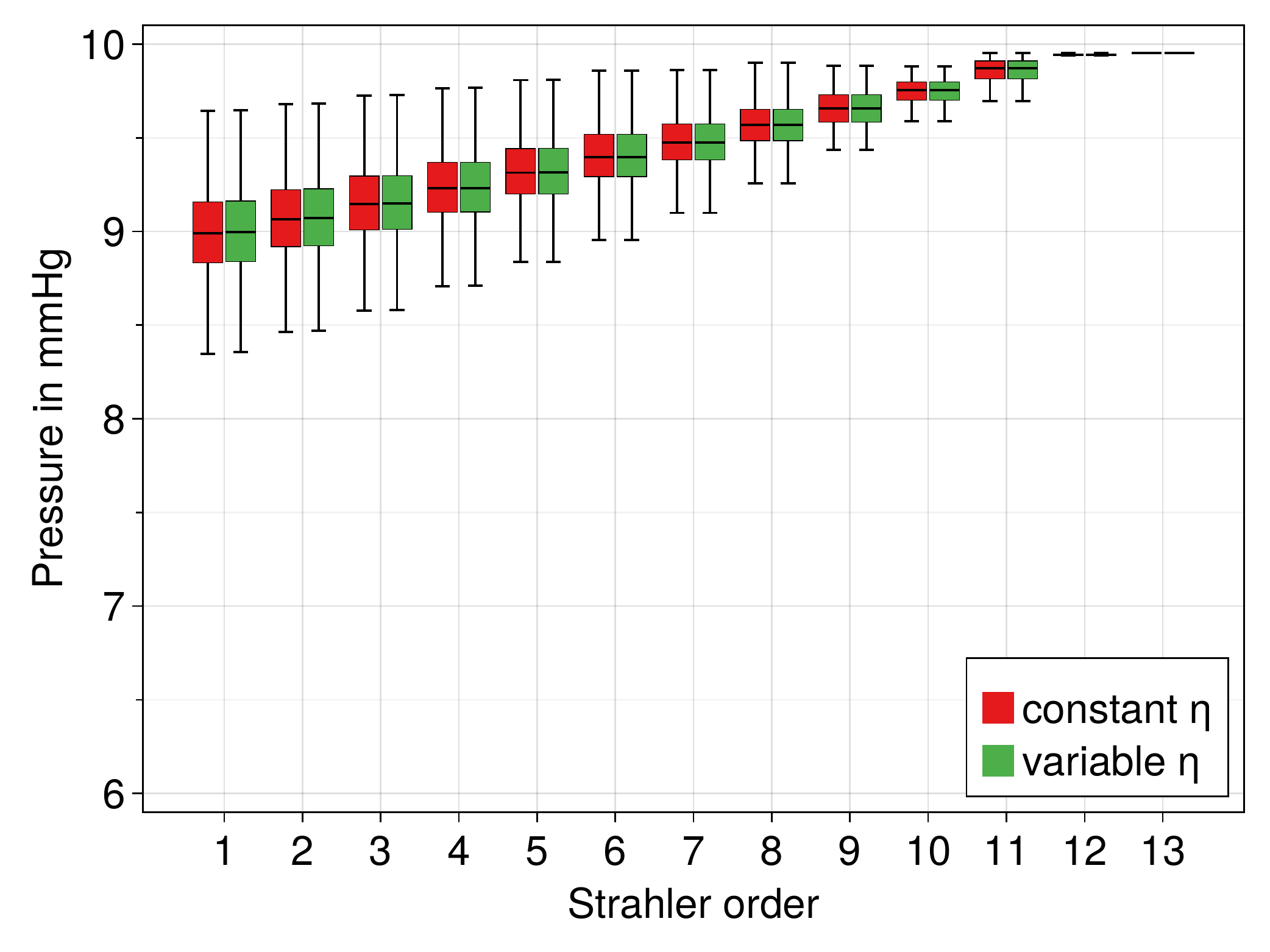}%
    \label{fig:pressure_min_power}}\hfill
  \subfloat[Volume minimization ($p\tsb{root} = \SI{10}{\mmHg}$)]
  {\includegraphics[trim=0 0 0 0,clip=true,width=0.5\textwidth]
  {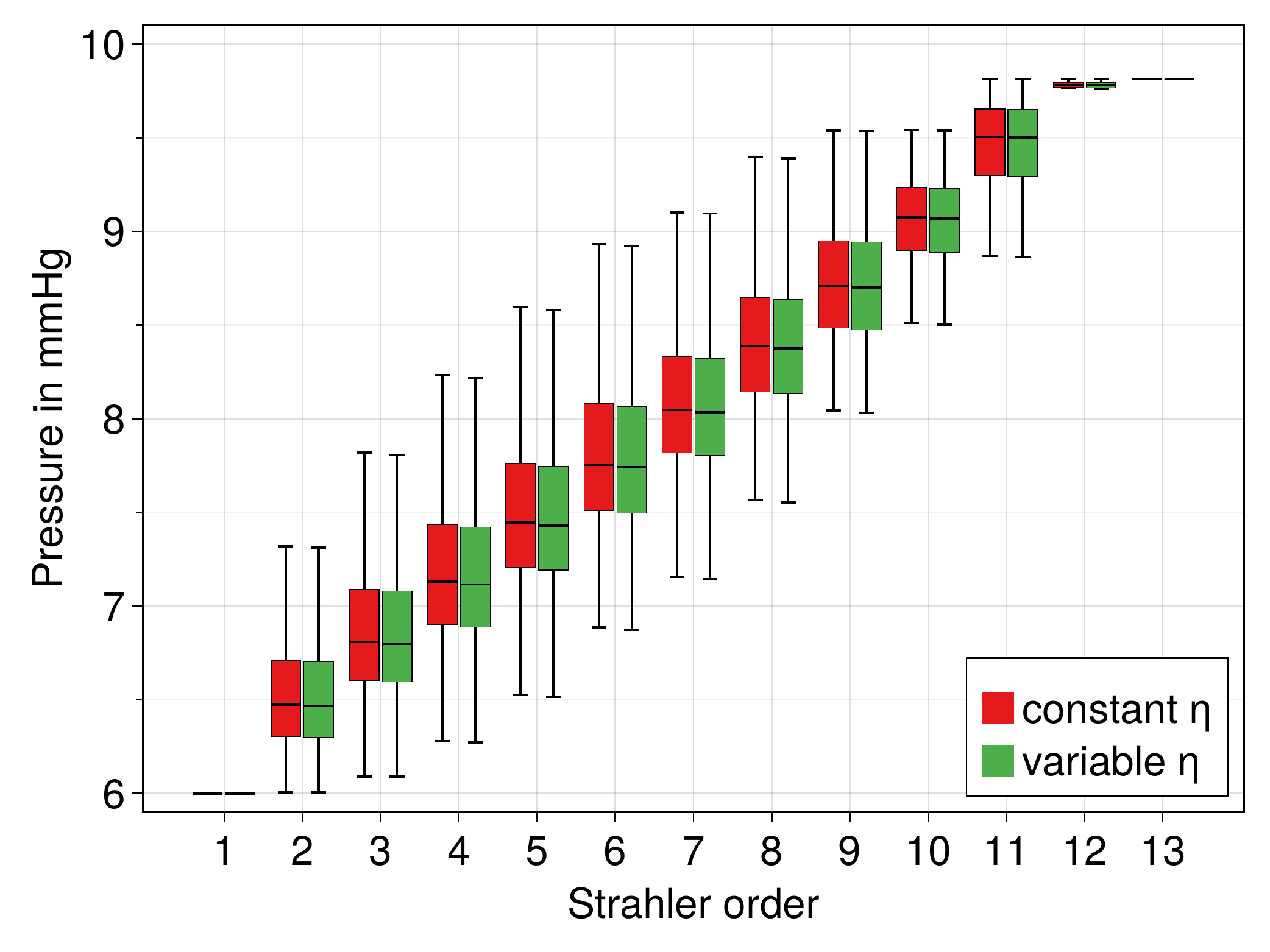}%
    \label{fig:pressure_min_volume}}
  \caption{Influence of Fåhræus--Lindqvist effect on pressure values of the distal nodes of vessels for different Strahler orders}
  \label{fig:pressure_comparison}
\end{figure*}

\subsection{Branching behavior}
The resulting branching exponents of all variants are summarized in row 5 of \cref{tab:results_overview}.
For "Power minimization" with constant viscosity $\etac$, the exponents are constant with $\gamma = 3.0$ across all branches regardless of metabolic demand $m_b$.
In contrast, the inclusion of the Fåhræus--Lindqvist effect leads to deviations from $3.0$, with branching exponents reaching a minimum of $2.9$.
For "Volume minimization," exponents are not constant even for constant blood viscosity.
Instead, most values fall between $2.0$ and $3.0$, with the smallest outliers having values of $1.43$.

For a more detailed comparison, the probability density function of the branching exponents for both optimization cases is compared in \cref{fig:branching_exponent_comparison}.
The influence of the Fåhræus--Lindqvist effect shifts most branching exponents from a constant $3.0$ to $2.9$ during "Power minimization" (\cref{fig:bexp-power}).
During "Volume minimization" with constant blood viscosity, most exponents are at $3.0$ (\cref{fig:bexp-volume} in red) and are shifted to $2.9$ when including the Fåhræus--Lindqvist effect (\cref{fig:bexp-volume} in green).

\cref{fig:branching-exponent-mean} highlights the distribution of mean branching exponents across different branch types.
Each cell $(i,j)$ corresponds to a branch with child segments of Strahler order $i$ and $j$.
In \cref{fig:branching-exponent-mean}(a) the effect of variable blood viscosity on "Power minimization" is depicted.
The branching exponents decrease if the Strahler order of either child decreases, leading to the smallest branching exponent of $2.9$ at branches with two terminal segments.
If both child segments have a Strahler order over $8$, the mean branching exponent is at its maximum of $3.0$.
The effect of enforcing equal terminal pressure is shown in \cref{fig:branching-exponent-mean}(b).
Here, the higher the difference between the Strahler orders of both children is, the smaller the branching exponent is.
Again, the smallest mean branching exponents are observed at branches connecting two terminal segments with a value of $2.76$.
\cref{fig:branching-exponent-mean}(c) shows the accumulated effect of both constraints with a minimum mean exponent of $2.7$, again at terminal branches.

\begin{figure*}[ht]
  \centering
  \subfloat[Power minimization ($m_b = \SI{0.1}{\uW\per\cubic\mm}$)]
  {\includegraphics[trim=0 0 0 0,clip=true,width=0.5\textwidth]
  {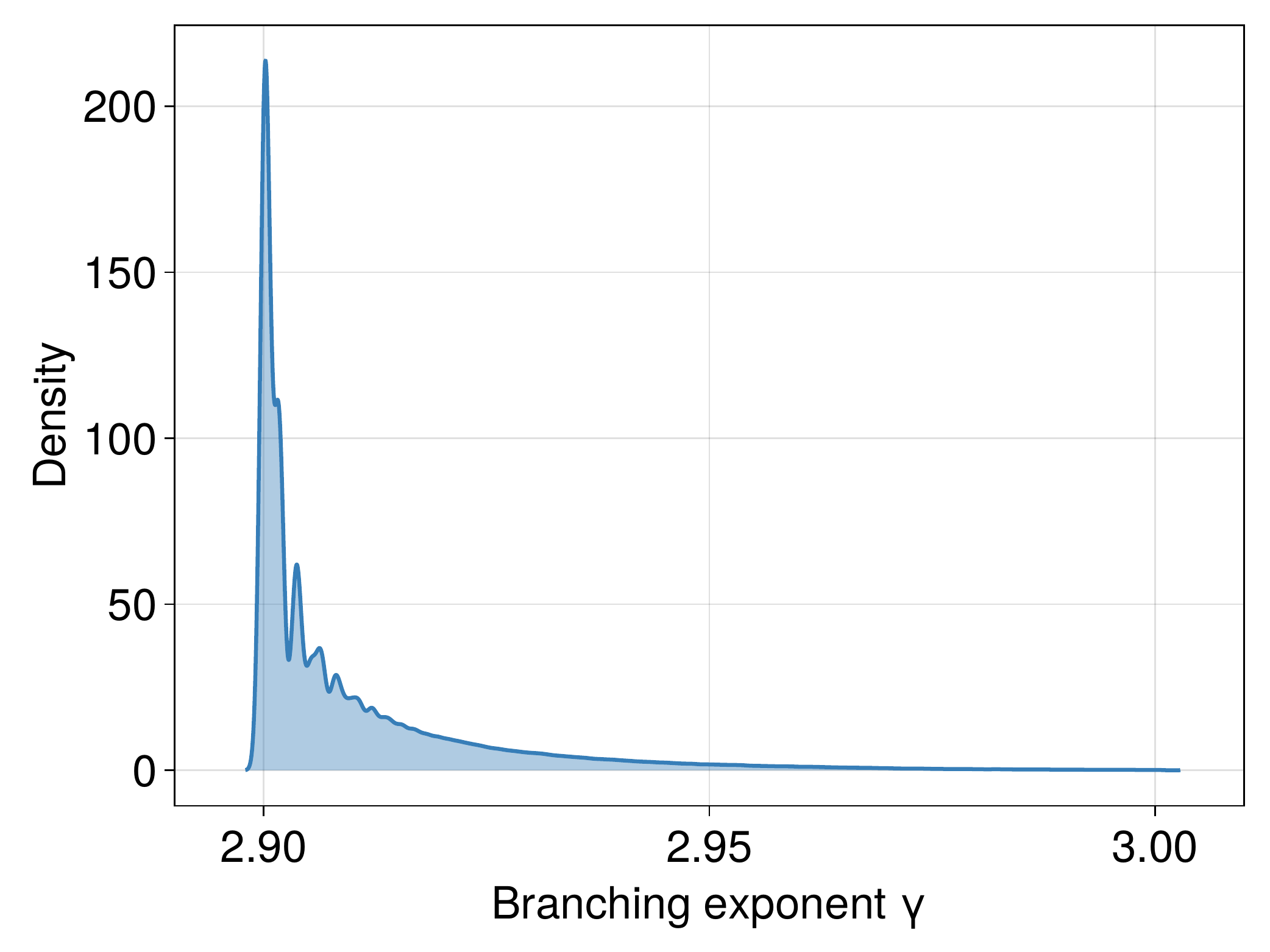}%
    \label{fig:bexp-power}}\hfill
  \subfloat[Volume minimization ($p\tsb{root} = \SI{10}{\mmHg}$)]
  {\includegraphics[trim=0 0 0 0,clip=true,width=0.5\textwidth]
  {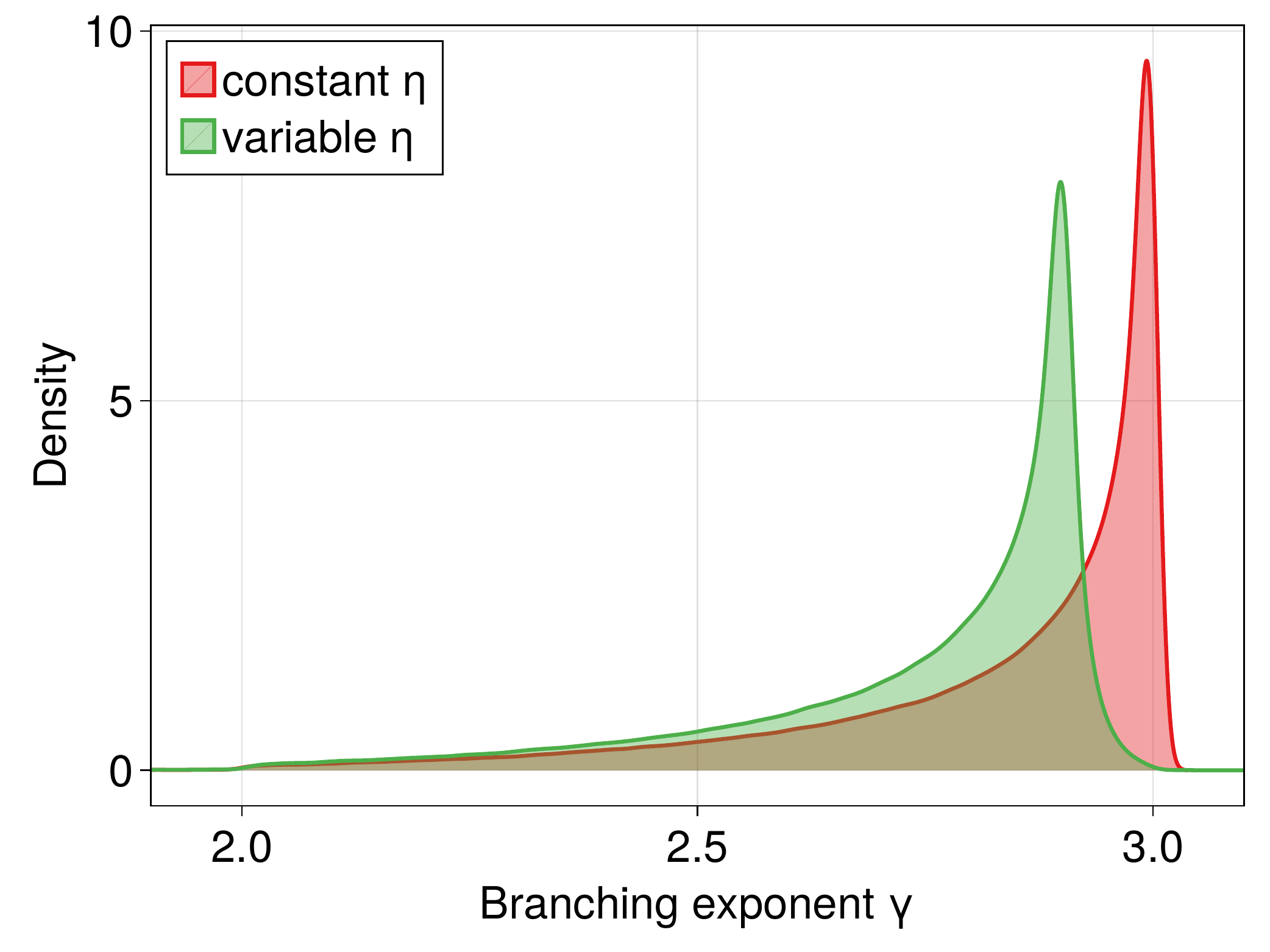}%
    \label{fig:bexp-volume}}
  \caption{Effect of enforcing variable viscosity and equal pressure onto the branching exponents}
  \label{fig:branching_exponent_comparison}
\end{figure*}

\begin{figure*}[ht]
  \centering
  \includegraphics[trim=0 0 0 0, clip=true,width=1.0\textwidth]{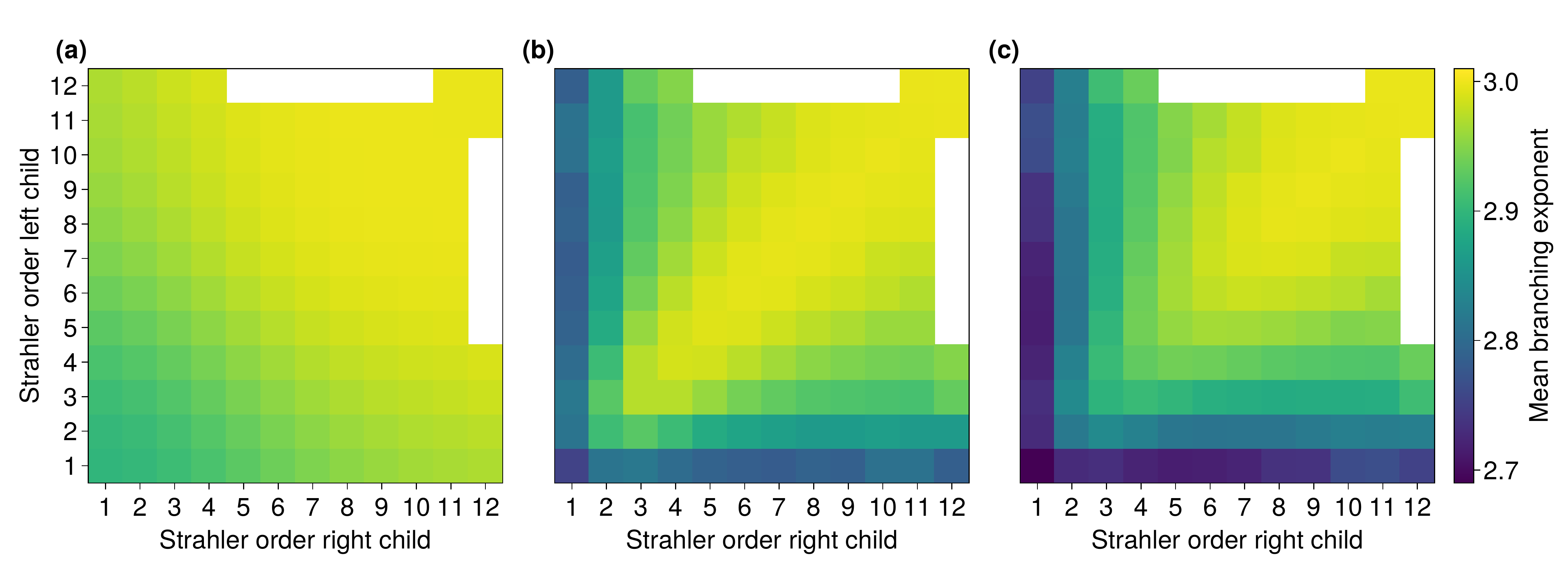}
  \caption{Mean values of branching exponent $\gamma$ for different types of branches, e.g., the cell $(1,2)$ corresponds to branches, where child vessels of Strahler order $1$ and $2$ meet. The results are symmetric. White cells correspond to branch types, which do not occur in the tree topology. \textbf{(a)} Power minimization with variable viscosity, \textbf{(b)} Volume minimization with constant viscosity, \textbf{(c)} Volume minimization with variable viscosity}
  \label{fig:branching-exponent-mean}
\end{figure*}

\subsection{Comparison to vascular corrosion cast}
We now directly compare our synthetic trees against a corrosion cast of the portal vein of the human liver \cite{debbaut2014analyzing}.
In \cref{fig:mean-radii}, the radii per generation (radius-adjusted Strahler order in reverse) are compared between the "Power minimization" with $m_b = \SI{0.1}{\uW\per\cubic\mm}$, the volume minimization with $p\tsb{root} = \SI{10}{\mmHg}$ and the corrosion cast data, including measurements and a best-fit trend line, based on the least sum of square errors.
The synthetic trees' radii fit the data and trend line well for the generations $5$ to $15$.
Notably, they significantly deviate for the first four generations, especially against the measurements with errors of around 25\%.
In contrast, the number of vessels in \cref{fig:number-of_vessels} of both synthetic trees fit the data of the corrosion cast well for all generations.
\begin{figure*}[ht]
  \centering
  \subfloat[Mean radii]
  {\includegraphics[trim=0 0 0 0,clip=true,width=0.5\textwidth]
  {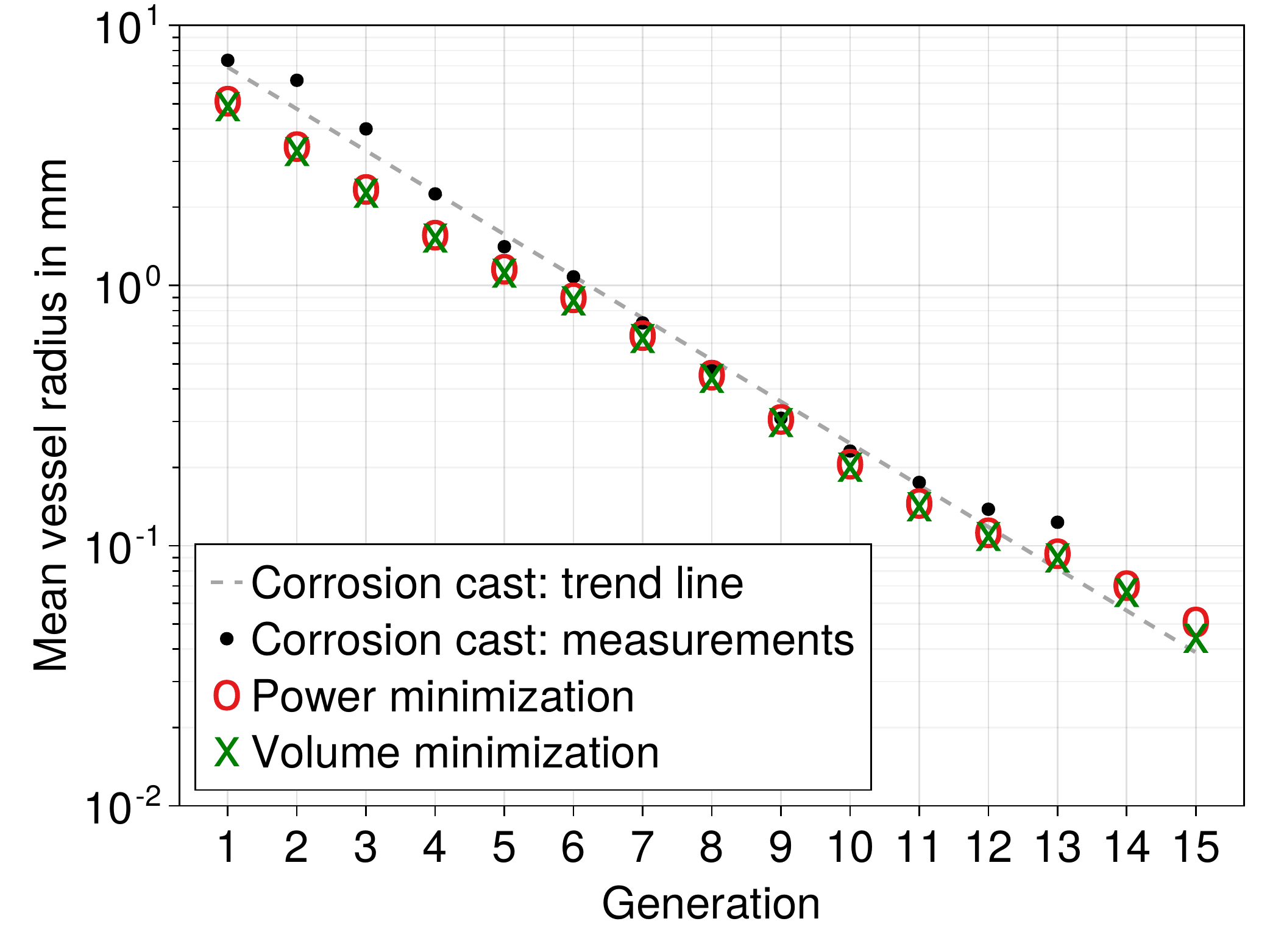}%
    \label{fig:mean-radii}}\hfill
  \subfloat[Number of vessels]
  {\includegraphics[trim=0 0 0 0,clip=true,width=0.5\textwidth]
  {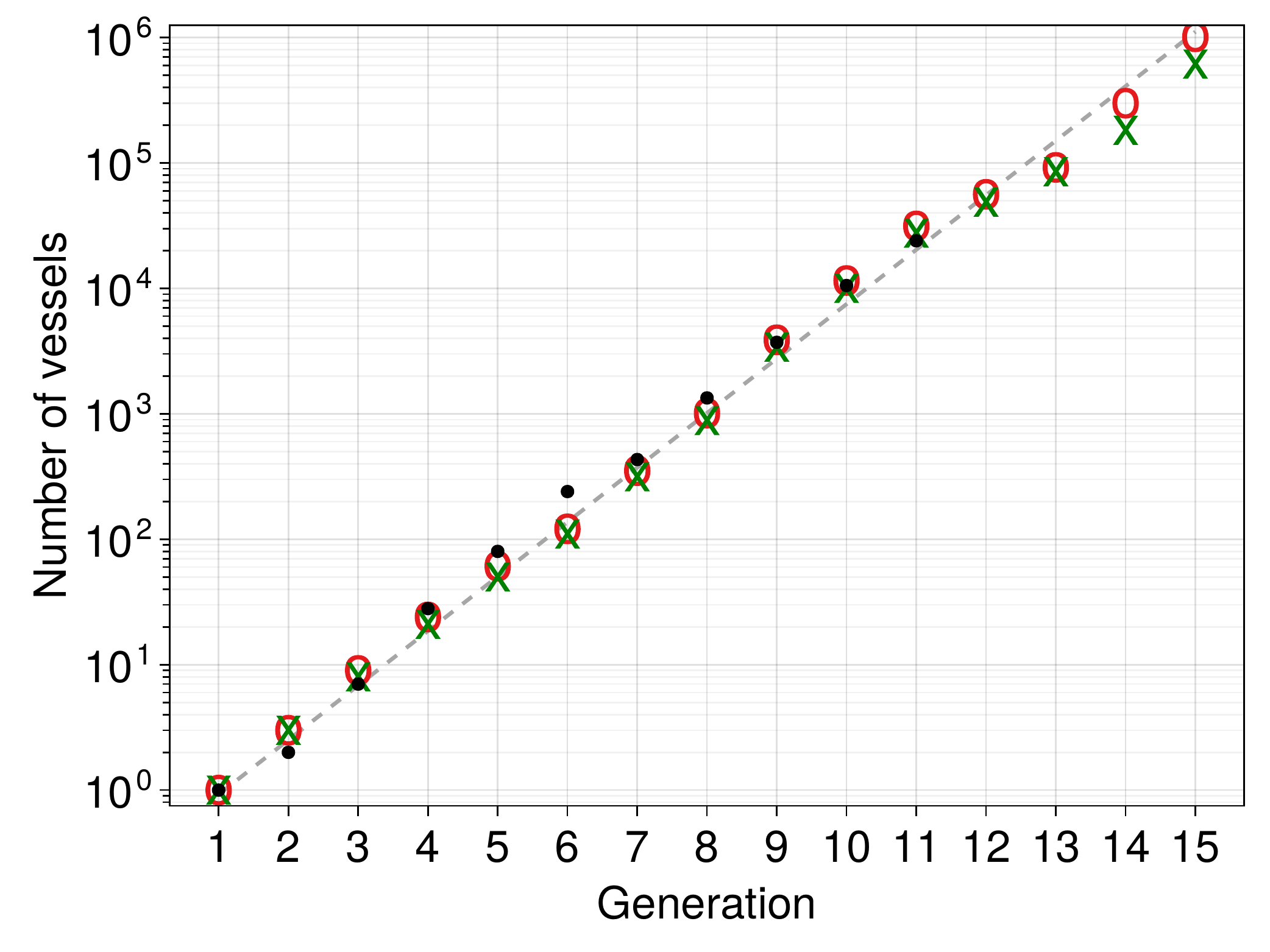}%
    \label{fig:number-of_vessels}}
  \caption{Comparison of synthetic trees (with variable viscosity) against corrosion cast data \cite{debbaut2014analyzing}}
  \label{fig:corrosion-comparison}
\end{figure*}

\section{Discussion}
The exclusion of Murray's law simplifies our two optimization problems and automatically allows the branching exponents to vary locally.
Both problems can generate synthetic trees with similar geometry and topology using appropriate parameters.
Power minimization with a metabolic factor of $m_b = \SI{0.1}{\uW\per\cubic\mm}$ leads to more realistic radii than a value of \SI{1.0}{\uW\per\cubic\mm}, which is in line with estimated values around $0.3-0.4$ \si{\uW\per\cubic\mm} for venous trees \cite{taber1998optimization, liu2007vascular}.
Similarly, a root pressure of $p\tsb{root} = \SI{14}{\mmHg}$ results in a pressure drop of $\Delta_p = \SI{8}{\mmHg}$, which is related to portal hypertension.
Such a high pressure drop leads to unrealistic small radii in comparison to a more realistic root pressure of $p\tsb{root} = \SI{10}{\mmHg}$.

Under power minimization, the radii are at their individual local optima,
i.e., the radius of each segment can be solved independently as the minimum of the metabolic demand and the viscous power dissipation.
This observation is in line with the findings of Murray and explains the constant branching exponent of $3$ in \cref{tab:results_overview}.
The variations in radii for Strahler orders $2$ to $12$ in \cref{fig:radius_min_power} are based entirely on the branching type, which is completely defined by the flow values of the child segments.

For volume minimization, no such simplification can be made, as the constraint of equal pressure creates dependencies between segments on the same path to the root. This constraint also forces radii to deviate from their local optima, which means a deviation from the branching exponent $\gamma = 3.0$.
The highest deviations are found at branches between two terminal segments, because they can be adjusted to a given pressure drop without significantly increasing the tree's volume.
Given the same length, they also constitute a higher pressure drop than segments with bigger radius.

The inclusion of the Fåhræus--Lindqvist effect reduces the viscosity of the smaller vessels and, in turn, increases their pressure drop.
The corresponding branching exponents also deviate from $\gamma = 3.0$, as can be observed in \cref{fig:branching-exponent-mean}.
In contrast, the effect on bigger vessels is negligible and results in constant exponents $\gamma = 3.0$, as seen in
\cref{fig:branching-exponent-mean}(a), for branches where both children have Strahler orders over $8$.

While the generated trees generally fit the corrosion cast data well, the underestimation of the largest radii (generation $1$ to $4$) is significant.
Since the vessels of the portal venous tree are more elliptical than circular, the radii of the vascular corrosion cast were estimated based on their cross-sectional area.
Perhaps the synthetic trees’ radii may correspond better with estimations based on other criteria such as a maximal inscribed circle. 
Another explanation is that the total energy dissipation is likely higher than our models predict.
This underestimation could be due to ignoring the effect of turbulent flow \cite{painter2006pulsatile} and simplified geometric modelling of the branching of vessels.
The effect of pulsatile flow, however, can be neglected because the blood comes not directly from the heart but first flows through the digestive tract, leaving only a limited amount of pulsatility inside the portal venous tree.

\section{Conclusion}
Our optimization framework can handle complex constraints and goal functions while generating synthetic trees up to but not including the capillary level of the microcirculation.
We used our framework to investigate the local branching behavior for different constraints and goal functions.
Branching exponents automatically lie in the experimentally predicted range between \num{2.0} and \num{3.0}.
Even small changes to Murray's original optimization problem, like the inclusion of variable blood viscosity, significantly affect the optimal branching exponents of vessels.
The topology and geometry of our synthetic trees closely follow the vascular corrosion cast of a human portal venous tree, with significant deviations only in the largest vessels.

Without enforcing any pressure constraint, terminal pressures vary by up to \SI{1.0}{\mmHg}, leading to highly heterogeneous boundary conditions for the microcirculation.
In contrast, enforcing equal terminal pressure in its current form leads to pressure variations up to \SI{2.0}{\mmHg} in the intermediate vessels of the mesocirculation.
Both these results need to be critically evaluated and compared against measurements of real vascular trees.
One approach to improve the pressure constraint could be to prescribe a certain physical range using inequality constraints rather than a fixed value.
In the future, we plan to include pulsatile and turbulent flow effects in a closed form, improving our framework's computational ability.
Furthermore, shear stress, a critical parameter for vascular growth, must also be incorporated into the model.

A more mature version of this model, specifically the ability to predict optimal branching exponents under different constraints, could have many potential applications in the medical field.
An example would be to predict and relate the branching behavior across the scales to vascular diseases.
These predictions could improve the interpretation of medical images by giving valuable input to the functional assessment of organs.

\section*{Acknowledgment}

The results presented in this paper were obtained as part of the ERC Starting Grant project ``ImageToSim'' that has received funding from the European Research Council (ERC) under the European Union’s Horizon 2020 research and innovation programme (Grant agreement No.~759001).
The authors gratefully acknowledge this support.

\bibliographystyle{ieeetr}
\bibliography{references}

\end{document}